\documentclass[aps,twocolumn,floatfix,groupedaddress]{revtex4}
\usepackage[dvips]{graphicx}
\usepackage{indentfirst}
\usepackage{amsmath}
\usepackage{times}

\begin{document}

\title{The Influence of Interatomic Bonding Potentials on Detonation Properties}

\author{Andrew J. Heim}

\author{Niels Gr{\o}nbech-Jensen}
\affiliation{Department of Applied Science,
University of California, Davis, California 95616}
\affiliation{Computational Research Division,
Lawrence Berkeley National Laboratory, Berkeley, California 94720}

\author{Timothy C. Germann}
\affiliation{Applied Physics Division, Los Alamos National Laboratory,
Los Alamos, New Mexico 87545}

\author{Edward M. Kober}

\author{Brad Lee Holian}

\author{Peter S. Lomdahl}
\affiliation{Theoretical Division,
Los Alamos National Laboratory, Los Alamos, New Mexico, 87545}

\date{\today}

\begin{abstract}
The dependence of macroscopic detonation properties of a two-dimensional diatomic (AB) 
molecular system on the fundamental properties of the molecule were investigated.  
This includes examining the 
detonation velocity, reaction zone thickness, and critical width as a function of the 
exothermicity ($Q$) of the gas-phase reaction 
($\textrm{AB} \rightarrow (1/2) (\textrm{A}_2 + \textrm{B}_2 )$) and  
the gas-phase
dissociation energy ($D_{e}^{\textrm{AB}}$) for 
$\textrm{AB} \rightarrow \textrm{A} + \textrm{B}$. 
Following previous work, molecular dynamics (MD) simulations with a reactive empirical 
bond-order (REBO) potential were used to characterize the shock-induced
response of a diatomic AB molecular solid, which exothermically reacts to produce 
$\textrm{A}_{2}$  and $\textrm{B}_{2}$ gaseous products. Non-equilibrium MD  
simulations reveal that there is a linear dependence between the square of the 
detonation velocity and each of these molecular parameters.  The detonation velocities 
were shown to be consistent with the Chapman-Jouguet (CJ) model, demonstrating 
that these dependencies arise from how the Equation of State (EOS) of the products 
and reactants are affected. Equilibrium MD simulations on microcanonical ensembles 
were used to determine the CJ states for varying $Q$s and radial 
distribution functions to characterize the atomic structure.  The character of this 
material near the CJ conditions was found to be rather unusual consisting of polyatomic 
clusters rather than discrete molecular species.  It was also found that there was a 
minimum value of $Q$ and a maximum value of $D_{e}^{\textrm{AB}}$ for which a 
pseudo-one dimensional detonation could not be sustained.  The reaction zone of this 
material was characterized under both equilibrium (CJ) and transient (underdriven) 
conditions.  The basic structure is consistent with the Zeldovich- 
von Neumann-D\"{o}ring model with a sharp shock rise and a reaction zone that 
extends to 200-300~\AA.  The underdriven systems show a build-up process which requires 
an extensive time to approach equilibrium conditions. 
The rate stick failure diameter (critical width in 2D) was also found to depend 
on $Q$ and $D_{e}^{\textrm{AB}}$. The dependence on $Q$ could be explained in terms of 
the reaction zone properties.
\end{abstract}

\maketitle


\section{Introduction}

The simplest theory of detonation is that of Chapman and Jouguet (CJ) 
\cite{Fickett, Mader, Davis1}. In this one dimensional 
theory the shock rise and reaction are treated as instantaneous. On a pressure-specific 
volume ($P$-$v$) state diagram the point of 
tangency between a Rayleigh line (an expression 
of the conservations of mass and momentum across the detonation front traveling at a given 
velocity) and a Hugoniot 
(conservation of energy) is the CJ state. The slope of the Rayleigh line is proportional to 
the negative of the square of the product of the initial density ($\rho _0$) 
of the material and the 
detonation velocity ($u_s$). If $u_s$  is any slower than the CJ value ($u_{sj}$), 
the Rayleigh line does not 
intersect the Hugoniot, and there is no solution to the conservation equations. In this 
light the CJ state is determined by the Equation of State (EOS) of the products and the 
initial state of the reactants, and that determines the minimum detonation velocity 
($=u_{sj}$) for the system. 
This hypothesis predicts the detonation 
properties of high performance high explosives (HE) reasonably well despite its crude assumption 
\cite{Fickett, Mader},

A more detailed model is the classical theory of detonation due to Zeldovich 
\cite{Zeldovich}, 
von Neumann \cite{vonNeumann}, and D\"{o}ring \cite{Doering} (ZND) 
which, following the initial shock compression, 
allows the molecules of a high explosive (HE) to react and expand. 
This is represented by a pressure profile, 
whose principal features are (1) an instantaneous shock rise to a state where the 
reactants are heated and compressed, and this is typically referred to as the von 
Neumann (VN) spike; 
(2) a fixed-width reaction zone, in which reactions provide the chemical energy to 
maintain the detonation wave as density and pressure decrease; and  
(3) a Taylor wave of the rarefying product gases. 
In the case where the detonation is supported by a driving piston, there will be a constant 
state in the pressure profile from some point behind the reaction zone back to the piston. 
If the piston is driven at 
the particle velocity of the unsupported final state (matching the CJ state at the end of 
the reaction zone),  there will be no Taylor wave, and only the reaction zone will be 
observed. If the piston is driven at a greater velocity than this critical value, then the 
detonation velocity will be increased. The detonation is now said to be overdriven, and 
the flow in the constant zone is subsonic in the frame of the detonation front. 
For the case where the detonation is underdriven with respect to the CJ conditions, it 
should asymptote to the minimum detonation velocity determined by the CJ state, and only the 
Taylor expansion will be affected by a disturbance behind the final state \cite{Fickett}. 

Molecular Dynamics (MD) simulations are well suited to test these and other aspects of 
detonation theory and their associated models under controlled microscopic conditions 
as demonstrated by work going back over a decade 
\cite{Brenner,White,Rice1,Rice2,Swanson,Maffre,Erpenbeck,Robertson,Haskins,Fellows,Germann,Holian,Maillet}.  With MD it is possible to control the inherent material properties, initial 
material state, and confinement conditions of the simulation.
For example, using a predissociative Morse potential, Maffre \emph{et al.} performed a 
preliminary study of ``hot spots'', which arise at heterogeneities.~\cite{Maffre}. 
Similar work was also done using voids and gaps as the heterogeneities \cite{Germann,Holian}.
Monte Carlo techniques have been incorporated with MD to find
thermodynamic properties and the Hugoniot of a system of hard spheres \cite{Erpenbeck} and, 
more recently, a reactive model close to the one used here \cite{Rice1}.
Tests of the dependence of the critical flyer plate velocity needed to initiate detonation  
on the flyer plate thickness \cite{Haskins} have been
studied. Rice \emph{et al.}\ have characterized some aspects of  
the reaction mechanism \cite{Rice2} and demonstrated correspondence to hydrodynamic 
theory for a model system \cite{Rice1}. 
Attempts have been made to connect the micro- and macroscales with 
MD and hydrodynamic codes \cite{Maillet}.

Several of these simulations 
\cite{Brenner,White,Rice1,Rice2,Swanson,Robertson,Haskins,Fellows,Germann,Holian,Maillet} 
have been conducted in two dimensions using
a Reactive Empirical Bond Order (REBO) potential \cite{Brenner}, representing a 
simple material composed of two atom types (A and B). REBO is a modification of 
Tersoff's Empirical Bond Order (EBO) potential \cite{Tersoff}.  The restriction to two 
dimensions allows significantly greater spatial and time scales to be accessed for given 
computational resources, though three dimensional simulations have also been pursued. The
process of shock-induced reversible chemical reactivity, converting reactant AB molecules 
exothermically into $\textrm{A}_{2}$ and $\textrm{B}_{2}$ products, is 
represented in the AB model of Brenner \emph{et al.} \cite{Brenner}. With this, it has been 
demonstrated that non-equlilibrium MD (NEMD) simulations using REBO potentials 
produce detonations consistent with 
continuum theory and experimental observations \cite{Swanson}.
However, most of these prior studies with REBO have focused on using a single set
of materials characteristics and energetic parameters.  It has been found that  seemingly 
subtle variations in the model or parameters can result in rather dramatic changes in 
behavior.  Also, because of computational requirements, the simulations have been limited 
in scale such that significant features are not always resolved.  Utilizing large-scale 
simulations done with the SPaSM parallel MD code \cite{Lomdahl}, 
our aim is to extend the work of
Haskins \emph{et al}. \cite{Fellows} by investigating the dependencies
of the detonation velocity, the reaction zone thickness, and the critical width that a HE 
must have in the transverse directions to sustain detonation on these atomistic energetics.  
This will be done with controlled variations in the
fundamental microscopic energetic quantities, namely
the exothermicity ($Q$) of the reaction and AB dissociation energy ($D_e ^{\textrm{AB}}$), 
in order to document more 
thoroughly the relationship between microscopic properties and macroscopic behavior.

This paper is organized as follows. 
In Sec.~\ref{sec:method} we lay out the details of the potential and the simulations used. 
In Sec.~\ref{sec:vel}, using equilibrium microcanonical (NVE) ensembles, 
we map out the product Hugoniot for set values of parameters and compare the
expected CJ velocities of the detonation fronts to those found by unsupported NEMD simulations. 
The CJ states of these materials are also characterized here.

In Sec.~\ref{sec:eos} we study the relationship of $Q$ to the equation of 
state (EOS) in order to better 
understand the results from Sec.~\ref{sec:vel}.  In Sec.~\ref{sec:crit} we 
characterize the width of the reaction zone and compare this to the critical minimum  
width ($W_{c}$) which a 2D sample must have in the direction normal 
to detonation front's propagation in order for that detonation to be sustained. 
(In 3D cylindrical samples, of course, the critical width analog 
is known as the failure diameter, and the experimental setup is called a rate stick.)


\section{Methods}
\label{sec:method}
There are several versions of REBO used in the related literature. The version used here is
due to Brenner \emph{et al.} \cite{Brenner}.
In it the binding energy of an N-atom system takes the form,
\begin{equation}\label{eq:REBO}
E_{b}=\sum_{i}^N\sum_{j>i}^N\{f_{c}(r_{ij})
[V_{R}(r_{ij})-\overline{B}_{ij}V_{A}(r_{ij})]+V_{\textrm{vdW}}(r_{ij})\},
\end{equation}
where $r_{ij}$ is the distance from atom $i$ to atom $j$. 
$V_{\textrm{A}}$ and $V_{\textrm{R}}$ are the attractive and repulsive terms, 
respectively, of a Morse intramolecular potential  $V_{\textrm{A}} - V_{\textrm{R}}$, 
and $\overline{B}_{ij}\equiv (B_{ij}+B_{ji})/2$, which
contains the effective valence and is designed to create dimers.
Depending on the local environment, $\overline{B}_{ij}$ varies from 0 to 1. 
If atom $i$ has no neighbors
(defined by $f_{c}(r)$) other than $j$, $B_{ij}=1$ and the full
Morse attraction is felt. On the other hand, if $i$ has two neighbors,
$j$ and $k$, and $r_{ij}<r_{ik}$, $0<B_{ik}<B_{ij}<1$, i.e., the
$ij$ and $ik$ attractions are both reduced, but more so for the pair
farther apart $(ik)$ than for the nearer pair $(ij)$. The effect of this
is to introduce a preferential valence of one for each atom.
A weak intermolecular van der Waals (Lennard-Jones form) 
interaction $V_{\textrm{vdW}}$ stabilizes a 
crystalline AB molecular solid, at least at low temperatures.  
$f_{c}$ is a
cutoff function which smoothly takes the potential to zero within a
finite distance.
The rest of the precise functional forms and parameters are contained in the 
errata of \cite{Brenner}.


Isolated XY molecules (X,Y~$\in$~\{A,B\}) have binding energies $D_e^{\textrm{XY}}$ 
(since $\overline{B}_{ij} = 1$); these are the fundamental parameters which we will vary from their 
baseline values \cite{Brenner}, $D_e^{\textrm{AA}}=D_e^{\textrm{BB}}=5.0$~eV and 
$D_e^{\textrm{AB}}=2.0$~eV\@.
These two parameters (constraining $D_e^{\textrm{AA}} = D_e^{\textrm{BB}}$) 
are related to the exothermicity ($Q$) through:
\begin{subequations}\label{eq:exo}
\begin{eqnarray}
\textrm{2AB}\rightarrow\textrm{A}_{2}+\textrm{B}_{2}+2Q\\
Q=D_{e}^{\textrm{AA}}-D_{e}^{\textrm{AB}}.
\end{eqnarray}
\end{subequations}
$D_e^{\textrm{AB}}$ is just the energy required to dissociate an isolated AB molecule:
\begin{equation}\label{eq:act}
\textrm{AB}+D_{e}^{\textrm{AB}}\rightarrow\textrm{A}+\textrm{B}.
\end{equation}
$Q$ is varied from 1.5~eV to 10.0~eV by holding 
$D_{e}^{\textrm{AB}}$ constant at 2.0~eV
while $D_{e}^{\textrm{AA}}$ is varied from 3.5~eV to 12.0~eV\@. $D_{e}^{\textrm{AB}}$ is 
varied from 0.5 to 3.5~eV, by varying $D_{e}^{\textrm{AB}}$ and $D_{e}^{\textrm{AA}}$ 
together so that their difference, $Q$, is constant at 3.0~eV\@.

\begin{figure}
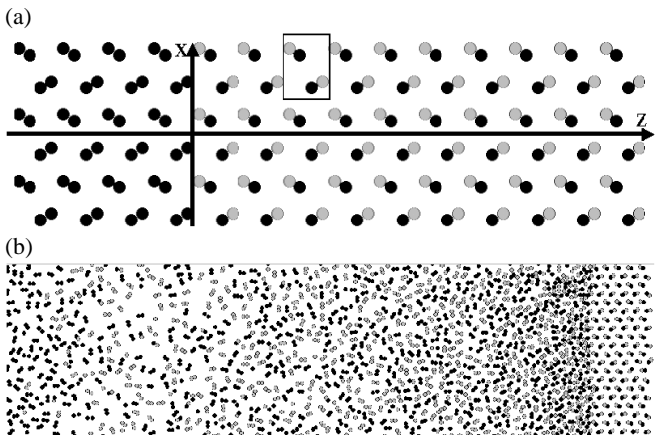

\raggedright 
(a)\\
\includegraphics*[width=\columnwidth]{img1_0.0000_Fig1a.epsi}
(b)\\
\includegraphics*[width=\columnwidth]{img1_0.0005_Fig1b.epsi}
\caption{(a) A snapshot of a magnified section of the initial sample in a flyer driven 
unsupported NEMD simulation of detonation at $t=0$. 
The A atoms are black 
and the B gray. The A$_2$ flyer plate can be seen to the left of the x axis. 
The box encloses a herringbone lattice cell, consisting of 2 AB dimers.\\
(b) A snapshot of the detonating sample with the detonation front moving toward the right.
\label{fig:ABic}}
\end{figure}

In each NEMD setup a stable A$_2$ flyer plate, four~cells thick (two A$_2$ dimers per cell), 
impacts a 2D metastable AB herringbone lattice at $z=0$ and a velocity of 9.8227~km/s
(Fig.~\ref{fig:ABic}). The sample has an initial temperature of 11.6~K. 
(The potential parameters roughly correspond to $\textrm{N}_2$, 
with a correspondingly low melting and boiling point.)
The resulting unsupported shock travels 
to the right, $\hat{z}>0$. In $\hat{z}$ the boundaries are culled, 
i.e., particles are lost when they exit the simulation cell. 
In the lateral $\hat{x}$ direction the boundaries are periodic in 
Sec.~\ref{sec:vel} to study planar detonations 
or culled, but padded with free space, 
in Sec.~\ref{sec:crit} to determine failure widths. 
In computations involving the reaction zone thickness and 
detonation velocity, the sample initially is at least 48~lattice cells in $\hat{x}$ and 600 
in $\hat{z}$. The simulation is allowed to run for at 
least 30.54~ps with a timestep $dt=0.25$~fs. The minimum duration is the same for trials 
determining $W_{c}$ versus ~either $Q$ or $D_{e}^{\textrm{AB}}$. The maximum 
length in cells for these calculations is 360~cells in $\hat{z}$\@.
In the equilibrium MD calculations presented in Sec.~\ref{sec:eos}, 
the samples are $25 \times 25$~$\textrm{cells}^{2}$ 
with periodic boundary conditions. The 
simulations are run for 40~ps with measurements averaged over the last 30~ps.

\section{Detonation Velocity and the CJ conditions}
\label{sec:vel}

\begin{figure}
\includegraphics*[width=\columnwidth]{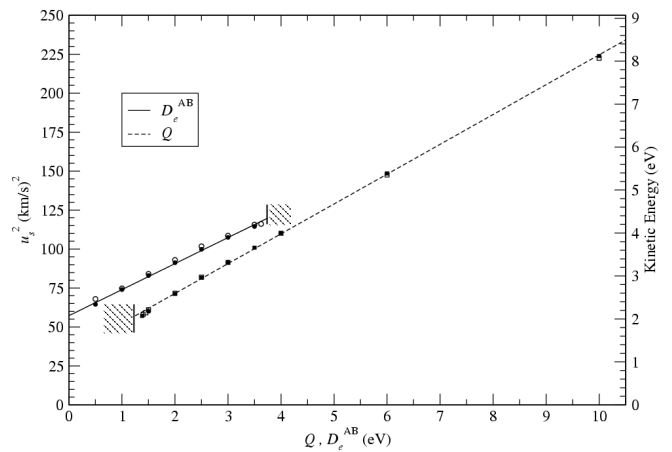}
\caption{Square of the detonation velocity ($u_s^2$) vs.~AB dissociation energy 
($D_{e}^{\textrm{AB}}$) 
and exothermicity ($Q$). The fl\mbox{}yer thickness and impact velocity are held 
fixed, so detonation is not sustained in regions where it is reported to in 
\cite{Fellows}. Failure is indicated by the shaded boxes. 
A sustained detonation is defined as not failing before the right edge of the sample.
Error bars are smaller than the size of the symbols. The scale on the right gives the 
kinetic energy of an atom traveling at $u_{s}$. The filled symbols are from NEMD simulations 
with a free boundary at $z=0$. The empty symbols are from NEMD simulations with a 
momentum mirror at $z=0$. The linear fits are $u_s^2 = 33.205 + 19.137 Q$ and 
$u_s^2 = 57.251 + 16.694 D_e^{\textrm{AB}}$. 
\label{fig:vel2bth}}
\end{figure}

Haskins \emph{et al}. report a linear dependence of  
$u_{s}^{2}$ on $Q$ for a similar AB material \cite{Fellows}. 
We repeat this study with a slightly different REBO potential and further it by varying the AB 
dissociation energy ($D_{e}^{\textrm{AB}}$). One expects the velocity to increase with $Q$ 
because the increased exothermicity of the reaction increases the temperature and pressure of 
the products. Fig.~\ref{fig:vel2bth} confirms the linear relationship between $u_{s}^{2}$ and 
$Q$\@ for this particular system. The differences between values in 
Fig.~\ref{fig:vel2bth} and \cite{Fellows} 
are due to differences in other REBO parameters and the flyer's thickness and 
impact velocity. Here, impact velocity is not raised above 9.8227~km/s in order to
avoid the ``fast detonation regime'' \cite{Maffre} (which may have been an 
artifact of the model used there).  At values of $Q < 1.5$~eV, the linear relationship 
begins to fail, and the system would not sustain a propagating detonation for $Q = 1.3$~eV\@. 
 This failure point likely arises because the reaction rate (determined by the temperature at 
the initial shock front) has become sufficiently slow such that it does not approach 
completion within the subsonic region of the reaction zone \cite{Fickett}.  The temperature 
at the shock front can be estimated from the kinetic energy of the shock front, given by the 
right-hand axis labels in Figure~\ref{fig:vel2bth}.

The dependence of $u_{s}^{2}$ on $D_{e}^{\textrm{AB}}$ is also found to be linear and 
increasing (Fig.~\ref{fig:vel2bth}).   Our initial expectation was that the variations 
in $D_{e}^{\textrm{AB}}$ would primarily affect the activation energy of the reaction and 
could quench the detonation when the activation energy became too great to be readily 
overcome at the temperature of the initial shock state.  The failure to maintain a 
propagating detonation for $D_{e}^{\textrm{AB}} > 3.7$~eV is possibly a manifestation 
of this.  The strong dependence of $u_{s}^{2}$ on $D_{e}^{\textrm{AB}}$ indicates that 
other aspects of the system are likely being affected by this perturbation.

To understand these relationships more thoroughly, we turn first to the 
basic test of standard detonation theory, which is to compare predictions based on the 
CJ state determined from equilibrium MD simulations with the evaluation of 
detonation propagation from NEMD studies.
The theoretical CJ state for a 1D detonation is at a sonic point at thermochemical equilibrium
\cite{Fickett}\@. If this is an improper assumption and the reaction is incomplete at the 
arrival time of the sonic point,
it could account for a discrepancy between the theoretical and actual detonation velocities. 
For a slightly different REBO potential, Rice \emph{et al.} found a detonation velocity 
from an unsupported simulation that was 6.1\% 
lower than that found from the calculation of their CJ state
\cite{Rice1}\@. 

Our procedure for locating the
CJ state is described here and similar to that of Rice \emph{et al.} \cite{Rice1}\@. 
At different values of specific volume ($v$), sets of microcanonical (NVE) MD simulations 
of 2500 particles are run for 40 ps, providing sufficient time to equilibrate. 
The onset of equilibrium is determined by the shape of the time evolution of the 
average properties of simulation. After each of these reaches a plateau, as determined 
by visual inspection, the simulation is allowed to continue. A runs-above-the-median 
test is performed to determine that the curves are flat with only random fluctuations. 
At each value of $v$, the value of specific internal energy ($E$) is sought that is 
a solution of the Hugoniot jump condition, 
\begin{equation}
\label{eq:Hug}
\frac{1}{2} (v_{0}-v)P = E-E_{0}
\end{equation}
(energy conservation in which $P_0 = 0$), where 
$P=\frac{1}{2}(P_{zz}+P_{xx})$ is the hydrostatic pressure, 
$P_{\alpha \beta}$ is the negative of the corresponding component of the stress tensor 
and has ideal and virial components, and the subscript 0 indicates the state in front of the 
detonation front. Once $\langle E \rangle$ is determined for the 
present value of $v$, NVE ensemble averages $\langle x \rangle$ of 
other thermodynamic quantities are computed by linear interpolation.

\begin{figure}
\includegraphics*[width=\columnwidth]{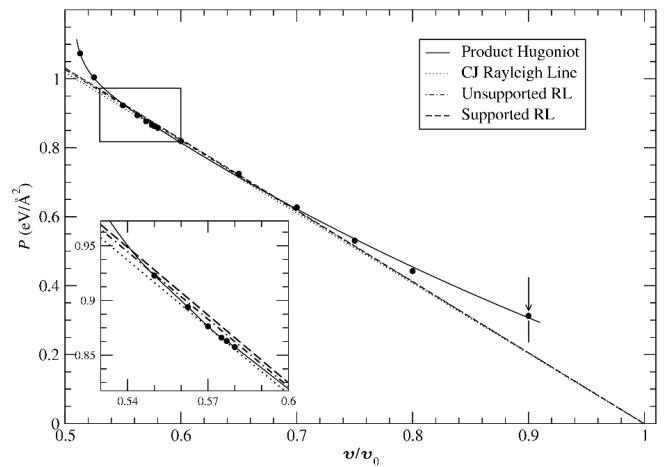}
\caption{A $P$-$v$ state diagram of the equilibrium product Hugoniot. 
The solid curve is a guide to the eye. The dotted/dashed curves are
Rayleigh lines plotted using the initial state ($P_{0}=0.0,v/v_{0}=1.0$) 
and a slope of 
-$u_{s}^{2}\rho_{0}$, where $\rho$ is the mass density and $u_{s}$ for each curve, from the  
steepest down, is the median value of shock velocity taken from the supported detonation, 
the EOS calculation, and the unsupported 
detonation. The arrows represent a series of simulations at constant $v$ used to find the 
datum to which they point. The box is a magnification.
\label{fig:Pv}}
\end{figure}

\begin{figure}
\includegraphics*[width=\columnwidth]{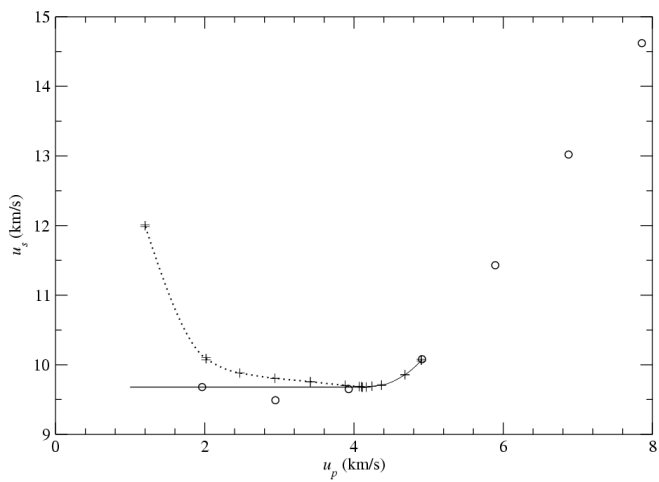}
\caption{Shock velocity ($u_{s}$) vs.~particle velocity ($u_{p}$). 
Similar plots can be made for other thermodynamic variables on the 
abscissa. Each point corresponds to a different value of specific volume ($v$), 
which increases along the abscissa in the direction opposite
that of $u_{p}$. A constant line is drawn at the determined minimum value of $u_{s}$. 
The dotted part of the curve indicates that which is not dynamically accessible. Instead 
NEMD simulations with pistons moving at $u_{p}$ less than the value at which the minimum is 
located ideally should follow the constant line. The open circles show the results of piston 
driven NEMD simulations where the piston is driven at $u_p$.
\label{fig:usup}}
\end{figure}

By repeating this procedure for several
values of $v$, the product $P$-$v$ Hugoniot is determined (Fig.~\ref{fig:Pv})\@.
Using the remaining jump conditions, mass conservation:
\begin{equation}\label{eq:cons_mass}
u_{p}/u_{s} = (v_{0}-v)/v_{0}
\end{equation}
and momentum conservation in which $P_{0}=0$,
\begin{equation}\label{eq:cons_momentum}
u_{s} u_{p} = v_{0} P,
\end{equation}
one can find $u_{s}$ as a function of the particle velocity at the final state 
($u_p$)\@. 
The minimum possible value of $u_{s}$ is the CJ value \cite{Fickett}, so the minimum of 
$u_{s}$
versus any thermodynamic parameter is at the CJ value of that parameter (e.g., see 
Fig.~\ref{fig:usup})\@. We use the minimum determined by these means as iterative 
approximations of the CJ state 
(see Table~\ref{tab:CJ} under the CJ-Interpolation column)\@. We refine the
process by fitting a quadratic through the points surrounding the minimum. 
We must be careful to include a domain small enough that a quadratic 
is a good approximation to the data, yet large enough to include more than three data points 
in order to get a proper estimate of the error from the goodness-of-fit parameter.

\begin{table*}
\caption{Values measured or defined in a process which determines the CJ state through a 
series of microcanonical equilibrium simulations 
or in the constant zone of a non-equlilibrium simulation of a critically 
supported detonation (Sec.~\ref{sec:eos}). $v$ is the specific volume. $u_s$ is the shock 
velocity. $u_p$ is the particle velocity at the final state. $T$ is the absolute 
temperature. $U$ is the potential energy. $E$ is the specific internal energy, and $P$ is 
the 2D pressure.  
Values marked with asterisks are input into the corresponding simulation(s).
The parentheses indicate the error in the last two digits of the corresponding reported 
value.}
\label{tab:CJ}
\begin{tabular}{|l|l|l|l|l|} \hline
& 				\multicolumn{3}{|c|}{CJ Interpolation}&			Supported NEMD 	\\ \hline
$Q$ (eV)& 			1.5*&			3.0*&		6.0*&		3.0*		\\ \hline	
$v/v_{0}$&			0.50182(24)&		0.57164(81)&	0.6171(19)&	0.5747(16)	\\ \hline
$u_{s}$ (km/s)&	    		8.0720(11)&		9.6758(90)&	12.2055(20)&	9.7360(40)	\\ \hline
$u_{p}$ (km/s)&			4.0212(13)&		4.1446(40)&	4.673(22)&	4.144*		\\ \hline
$\langle k_{B}T\rangle$ (eV)&	0.4017(14)&		0.7839(43)&	1.7550(90)&	0.7791(56)	\\ \hline
$\langle U\rangle$ (eV)&	-0.25156(79)&		-0.5605(21)&	-1.19657(90)&	-0.5746(72)	\\ \hline
$\langle E \rangle$ (eV)&	0.15012(65)&		0.2234(22)&	0.5584(84)&	0.2234306(15)	\\ \hline
$P$ (eV/\AA$^{2}$)&		0.70591(33)&		0.8726(13)&	1.2411(62)&	0.8541(83)	\\ \hline
\end{tabular}
\end{table*}

To test the CJ results found from these NVE simulations, 
we run a supported detonation simulation 
in which an infinitely massive driving piston impacts the AB sample at the CJ-determined 
velocity ($u_{pj}$). 
This should establish a
constant zone behind the front that should be at the CJ state, with the front propagating 
at the CJ conditions. Measurements are taken from this run and can be found in 
Table~\ref{tab:CJ} under the Supported-NEMD column. 
From Table~\ref{tab:CJ} one can see that many of the values fall within error of one another, 
although there are some slight discrepancies.  It should be noted that the supported-NEMD 
simulations can include transients from the initiation and/or build-up processes \cite{Mader}, 
and these will be explored further below.  Still, this discrepancy of 0.35\% difference from 
the unsupported detonation of $u_s = 9.70961\pm 0.00054$~km/s is quite 
small, especially since previous evaluations of this same system gave larger discrepancies
(9.3~km/s in \cite{Brenner} and 9.5~km/s in \cite{White}).  This is also 
less of an error than Rice \emph{et al.} found for their variation. 

\begin{figure}
\includegraphics*[width=\columnwidth]{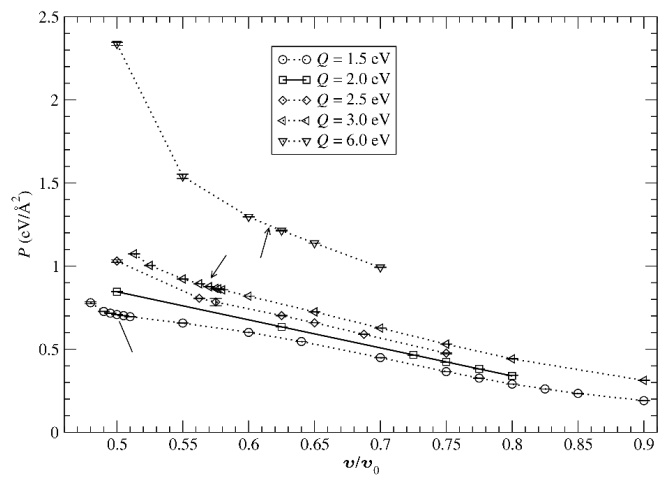}
\caption{Equilibrium Hugoniots for several values of $Q$. The CJ state is determined for the 
cases in which $Q$~=~6.0, 3.0, and 1.5 eV\@. The curve for $Q=2.0$~eV is well fit by a 
straight line of the form $P = 1.69 (1-v/v_0 )$. 
The dotted lines are all guides to the eye. As $Q$ decreases, the Hugoniots go from positive 
curvature to a zero curvature for $Q = 2.0$. 
The curve for $Q=1.5$~eV contains a section of negative curvature, indicating a phase 
transition. The arrows point to the determined CJ states. 
\label{fig:Hugs}}
\end{figure}

We repeated several of the tasks performed on $Q$~=~3.0~eV for several more values of $Q$\@. In 
Fig.~\ref{fig:Hugs} the equilibrium $P$-$v$ Hugoniots for several values of $Q$ are shown\@. 
Notice for $Q$~=~6.0~eV that the curve has a nice hyperbolic shape, for which the CJ state for 
this $Q$ is easy to determine. From it, as was done above, we find  $u_{sj}$ to be $12.20552 \pm 
0.0020$~km/s. The unsupported NEMD simulation with the free boundary gives 
$12.2031 \pm 0.0017$~km/s, a difference that is smaller than the error bars. 
As $Q$ decreases, the 
equilibrium Hugoniots flatten out, and it becomes more difficult to identify the exact 
point of tangency. At $Q$~=~2.0~eV the Hugoniot is well represented by a linear fit. 
The Hugoniot for $Q$~=~1.5~eV has a negative curvature (convex) 
section indicating the presence of some sort of phase 
transition (Fig.~\ref{fig:Hugs}). It should be noted that $Q$~=~1.5~eV is close to $Q_c$, the 
value for which a detonation can not be sustained. 

A comparison of the predicted and measured values of $u_s$ for different values of $Q$ are 
given Table~\ref{tab:Q}.  It is found that there is very good agreement between these values, with 
no discernible discrepancy for $Q$ = 6.0 eV, and only a moderate discrepancy of 2\% 
for $Q$ = 1.5 eV which is very near the failure point.  The latter point supports the idea 
that failure is occurring because of inadequate completion of the reaction when the 
sonic point passes.  Overall though, it can be concluded that these 
systems are responding in accord with the simple predictions of the Chapman-Jouguet 
hypothesis.  The variations in $u_s$ with the molecular parameters is therefore determined 
by how these effect the Equation of State (EOS) of the products.  These aspects will 
now be examined in further detail.

\begin{table}
\caption{As $Q$ is increased, the predicted ($u_{sj}$) and measured values ($u_s$) of the 
detonation velocity agree more. It is hypothesized that the increased exothermicity causes 
the reaction to come closer to completion by the sonic point.}
\label{tab:Q}
\begin{tabular}{|l|l|l|l|} \hline
  $Q$ (eV)& $u_{sj}$ (km/s)&$u_s$ unsupported&\% difference\\ \hline
  1.5&8.0720(11)&7.913(16)&$\approx 2.0$\\ \hline
  3.0&   9.6758(90)&9.70961(54)&$\approx 0.35$\\ \hline
  6.0&12.2055(20)&12.2032(17)&$\approx 0.0$\\ \hline
\end{tabular}
\end{table}


The curves in Figs.~\ref{fig:Pv} and \ref{fig:usup} are rather unusual. For a material 
behaving like an ideal gas, the former would be expected to have a 
hyperbolic shape and for the latter a parabolic shape. Here, for large 
$Q$ (= 6 eV) the resulting curves do have that type of behavior. However, as $Q$ is reduced, 
the Hugoniot flattens until it eventually gains a convex section, and the $u_s$-$u_p$ plot 
evolves into a curve with a double minimum.
Therefore we examine the CJ states more closely by simulating a microcanonical ensemble at 
the determined CJ values of density and internal energy. 

\begin{figure}
\includegraphics*[width=\columnwidth]{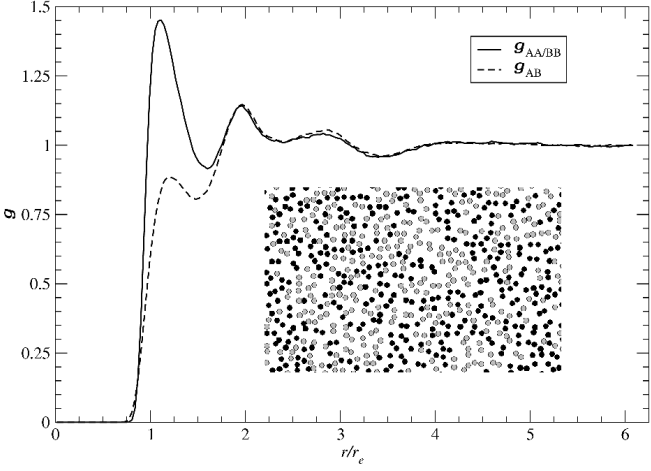}
\caption{Radial distribution function for the CJ state for $Q=3.0$~eV\@. 
$g_{\textrm{AB}}$ measures the chance of finding 
particles of 
opposite types a distance $r$ apart divided by the probability if the atoms were randomly 
distributed. $g_{\textrm{AA/BB}}$ indicates the probability of finding 
particles of the 
same type a distance $r$ apart, again, divided by the probability if the distribution were 
random. The inset shows a section of a snapshot of the simulation. 
A atoms are black and B are gray.
\label{fig:rdf}}
\end{figure}

A snapshot 
of the NVE-at-CJ simulation for $Q$ = 3.0 eV is shown in Fig.~\ref{fig:rdf},
along with the corresponding radial 
distribution functions (RDF), $g(r)$, for particles of the same and different type. It 
is evident from Fig.~\ref{fig:rdf} that, at the CJ state, the system is dominated by short AA 
and BB contacts, which would be expected for the product species.  The 
interatomic 
distances for these ($\approx 1.1~r_e$) is slightly larger than that defined for the 
isolated molecules 
($1.0~r_e$).  By the form of the potential, the introduction of a third particle within the 
defined bonding distance weakens the attractive part of the potential and moves the minimum 
of the bonding potential out to greater distances.  This aspect is readily explained by the 
high 
density at CJ\@. 

What is more intriguing is that there is also a significant number of close A-B interactions at 
$\approx 1.2~r_e$, showing that the system has not evolved into a simple mixture of A$_2$ and 
B$_2$. Also, there is a second peak in both RDFs at around $r/r_e = 2.0$.  
This second peak indicates that there 
are clusters of atoms forming at this compression. Looking at the snapshot embedded in 
Fig.~\ref{fig:rdf}, one can see that there appears to be linear oligomers forming. 

\begin{figure}
\includegraphics*[width=\columnwidth]{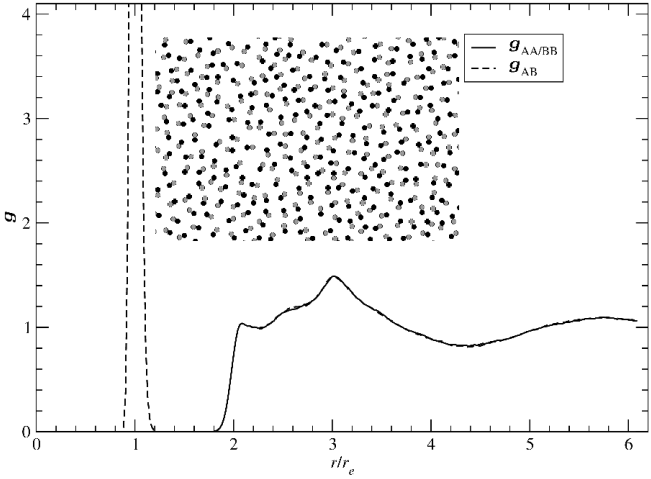}
\caption{RDF for melted AB. The first peak, the maximum of which is about 14, indicates the 
presence of dimers. The peak above $r/r_e = 2$ is due to the inner cutoff spline in the 
$V_{vdW}$ term of Eq.~\ref{eq:REBO}. The peak at $r/r_e = 3$ is probably caused by the 
arrangement of dimers.
\label{fig:rdf_melt}}
\end{figure}

For comparison Fig.~\ref{fig:rdf_melt} shows the RDF of an NVE 
simulation of AB at a temperature above its melting point and at the initial density, 
$v=v_0$. It has a strong peak for the AB dimers at $r/r_e = 1$ as expected.  Beyond this, 
it was expected that there would be a broad peak at ~2.8 $r_e$, which would be the van der 
Waals minimum.  Although this was observed, somewhat surprisingly, there is a sharp peak just 
above $r/r_e = 2$ which is present for all three atomic combinations, as well as another 
peak at 3 $r_e$. The former peak is caused by the positive slope of a section of 
the inner cutoff spline for the $V_{vdW}$ term, which creates an additional minimum in the 
multibody potential.  Examination of the inset show that the molecules still tend to line 
up, although the molecules are clearly separate diatomic species.
The peak at $r/r_e = 3$ is probably a result of this alignment and is the distance from a 
bonded partner of one atom to the non-bonded neighbor which is at ~2 $r_e$. These features 
highlight some of the unusual aspects that can arise from these complex interaction 
potentials.

The notable difference between the RDFs is that the one for the melted AB 
goes to zero between $r/r_e = $ 1.2 and 1.8, which highlights the clear diatomic nature of 
that system.  For the CJ state simulation, there is substantial intensity all across that 
region, and the RDF value barely drops below 1. 
Since there are no dissociated atoms or oligomers in the melted AB, it is reasonable that 
we see a domain above $r/r_e = 1$ in which a particle will not have a neighbor. 
Any third particle and its bound partner will be repelled 
by virtue of the $V_A$ term if they approach much closer than 2 $r_e$ to a particle in 
another dimer. At the CJ state density, this repulsion breaks down and clusters of particles 
form.  Similar behavior has also been observed in the systems that Rice \emph{et al.} 
studied, where describing the CJ state as one of atomization is a fair description.
\cite{Rice1,Rice2}.

\begin{figure}
\includegraphics*[width=\columnwidth]{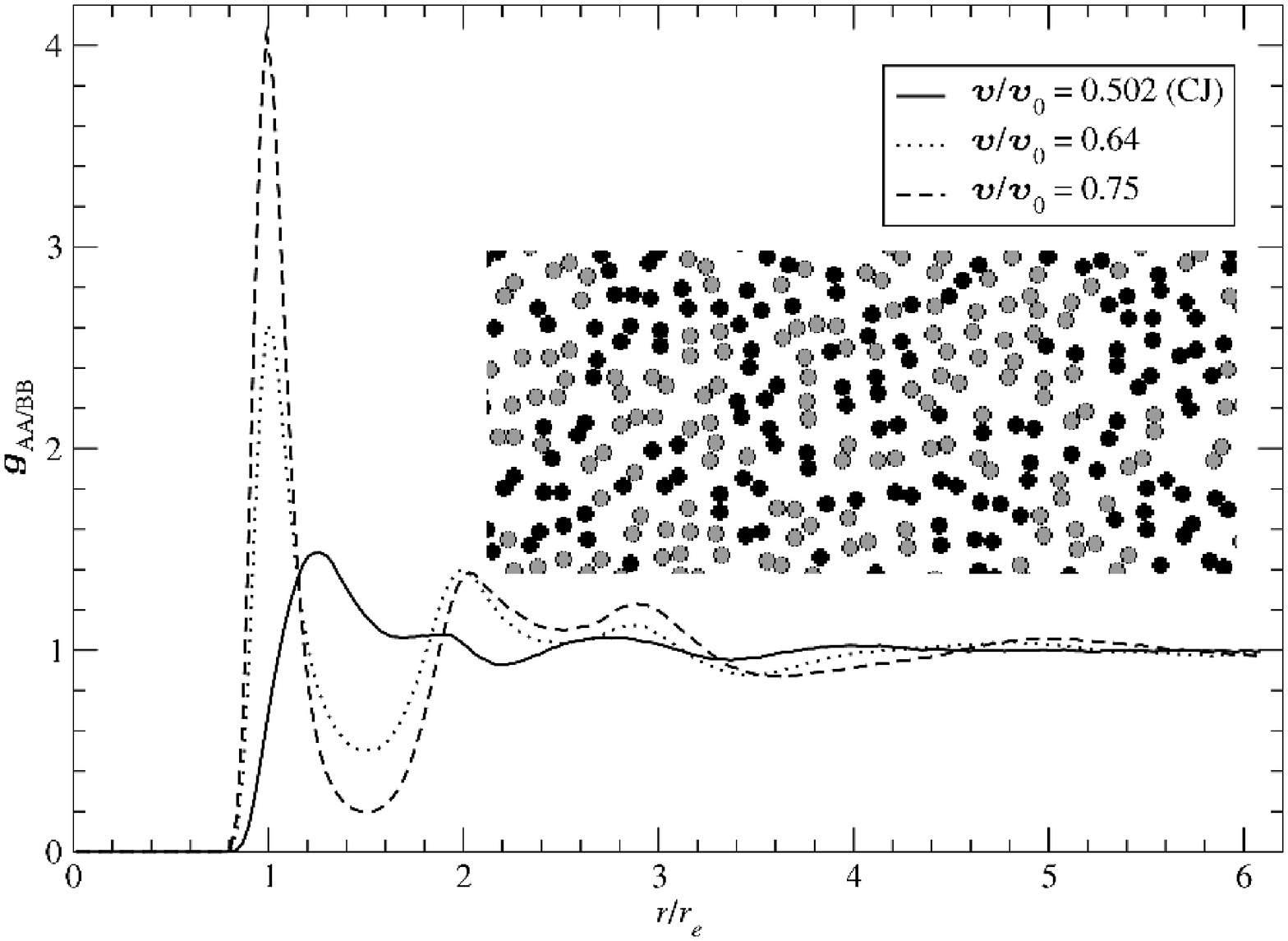}
\caption{Same-type radial distribution function for an exothermicity $Q =1.5$~eV 
at states along the Hugoniot at volumes shown. Notice the 
maximum at radial distance $r/r_e \approx 2.75$ for the CJ state. 
It represents the van der Waals 
equilibrium distance. The local minimum of the other curves at that point are still greater 
than unity. Notice the local maxima for the other two curves at $r/r_e =3.0$. They suggest  
tetramer formation or dimer alignment as with the AB melt. 
The inset is a snapshot of the simulation for $v/v_0 =0.64$.
\label{fig:RDFs}}
\end{figure}

To examine this behavior more closely, we examine the RDFs along the Hugoniot.  
In particular, we consider the case of $Q = 1.5$~eV which has a convex section that 
indicates a phase transition. The RDFs for several states along that 
Hugoniot including $v/v_0 = 0.64$, the volume at which the phase 
transition occurs, are shown in the Fig.~\ref{fig:RDFs}. 
For the CJ state, $v/v_0 = .502$, the RDF is 
similar to that for the CJ state at $Q = 3.0$~eV\@.  
The first peak occurs at ~1.2 $r_e$ with 
no strong minimums occurring at longer distances.  At $v/v_0 =0.64$, a deep minimum 
develops at $\approx 1.5~r_e$ along with peaks at $r/r_e = 2.0$~and~3.0. 
From the snapshot at $v/v_0 = 0.64$ (See inset.), one may be convinced that both 
distinct dimers and extended oligomers are present.  At $v/v_0 = 0.75$, the peak 
sharpening and minimum development are more distinct, and the profile is more reminiscent 
of the AB melt illustrated in Fig.~\ref{fig:rdf_melt}. 

\begin{figure}
\includegraphics*[width=\columnwidth]{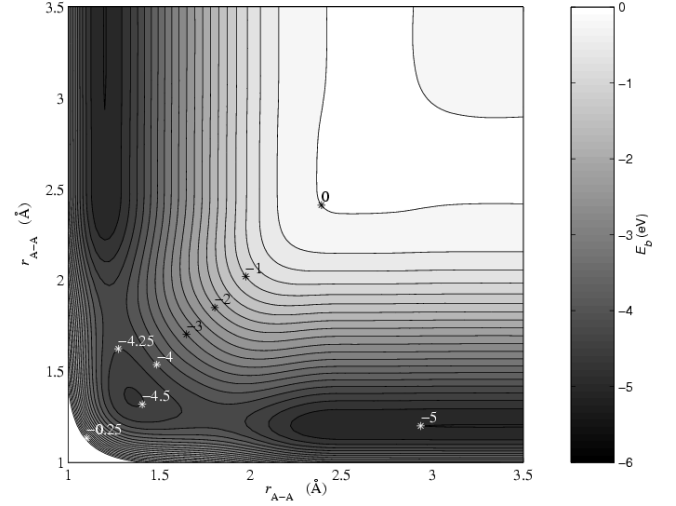}
\caption{Potential energy in eV vs.~separation distances in \AA~of three inline A (or B)
 atoms. The ``*'' marks the contour labeled by the nearby number. Notice the local minimum 
at (1.3, 1.3). It allows for trimer formation. No positive contours are shows, as would fill 
in the section near (1,1). The case for inline A-B-A (or B-A-B) looks similar yet shallower. 
\label{fig:AAA}}
\end{figure}

To investigate the cause of these clusters, we 
plot potential energy surfaces (PES) for the interatomic distances of three inline 
particles as Rice 
\emph{et al.} \cite{Rice2} did for a different version of REBO\@. 
From Fig.~\ref{fig:AAA}, which represents an A-A-A configuration, one can see that, 
at a distance of about 1.35~\AA, 
there is a minimum that allows for trimer formation. 
This configuration does 
not exhaust the possible configurations that may occur during reaction because it is 
constrained to be a linear conformation.  The interaction with neighboring atoms, 
particularly at CJ-type conditions, is also ignored and this could alter the absolute value 
of the activation energy ($E_a$) significantly.  However, as Table~\ref{tab:EavDeAB} 
suggests, $E_a$ seems to increase monotonically with $D_e ^{\textrm{AB}}$. This 
would rationalize that as $D_e ^{\textrm{AB}}$ is increased the material eventually fails to 
detonate given the same initiator strength.

\begin{table}
\caption{The minimum peak value of the barrier to be overcome when converting AB to BB or AA 
in 1 dimension ($E_{a,0}$, where the second subscript indicates the angle between AB and BB) 
for several values of $D_e ^{\textrm{AB}}$. Detonation 
cannot be initiated at $D_e ^{\textrm{AB}}=4.0$~eV for the given flyer thickness and 
velocity. All of the measurements of $E_{a,0}$ have an error of $\pm 0.0125$~eV\@.}
\label{tab:EavDeAB}
\begin{tabular}{|l|l|} \hline
  $D_e ^{\textrm{AB}}$ (eV)&$E_{a,0}$\\ \hline
  3.0&0.1875\\ \hline
  3.25&0.2375\\ \hline
  3.5&0.2625\\ \hline
  3.75&0.2875\\ \hline
  4.0&0.3375\\ \hline
\end{tabular}
\end{table}

\begin{figure}
\includegraphics*[width=\columnwidth]{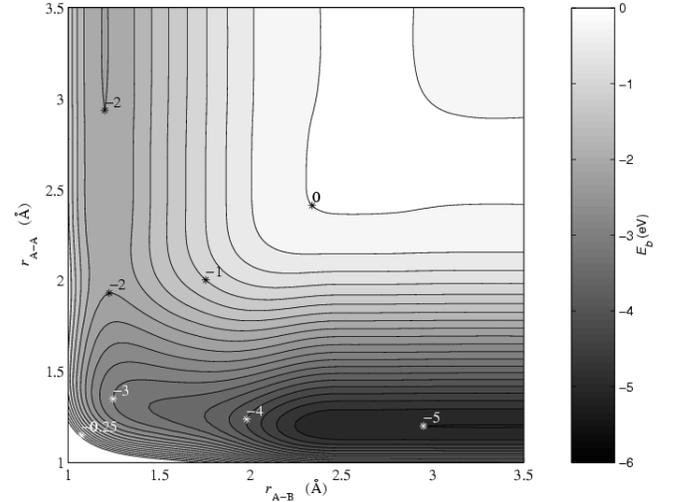}
\caption{Potential energy vs.~separation distances of an A-A-B (or B-B-A) configuration.
\label{fig:AAB}}
\end{figure}

\section{Exothermicity's Relationship to EOS}
\label{sec:eos}

We now turn to understanding the linear relationship between $u_s^2$ and $Q$.  One simple 
model is suggested by Fickett \emph{et al.} \cite{Fickett}, who derive a linear relationship 
between $Q$ and $u_s ^2$ by adding $Q$ to the incomplete equation of state of the 
initial state of a polytropic gas, an ideal gas with a constant specific heat.
An expression for the change in specific internal energy becomes 
$E-E_0 = (Pv-P_0 v_0 )/(\gamma -1)-Q$, where $\gamma$ is the adiabatic gamma 
\cite{Davis3}.
One can then eliminate $E-E_0$ with the Hugoniot jump condition (Eq.~\ref{eq:Hug}), 
thus solving for $P$. It is assumed that $P_0 = 0$. 
One can use the condition that Rayleigh line is tangent to the Hugoniot at the CJ state 
to arrive at 
$u_s ^2 = 2(\gamma ^2 -1)Q/m$, where $m$ is the mass of the reactants. If it is assumed that 
this EOS accurately describes our potential, $\gamma \approx 2$. 
A typical conventional HE has a $\gamma \approx 3$ \cite{Mader}.  
Since the volume at the CJ state is given by the expression 
$v/v_0 = \gamma /(\gamma + 1)$, this relationship does rationalize the somewhat greater
compression observed here ($v/v_0 < 0.67$) compared to that for conventional HE's 
($v/v_0 \approx 0.75$).  (It should be noted that the values of $v/v_0$ given in 
Table~\ref{tab:CJ} are smaller yet and imply a value of $\gamma \approx 1$.) However, 
as we have demonstrated that the product's EOS does not behave like an ideal gas, this 
does not serve as a good basis to explain the observed relationship between $u_s^2$ and $Q$.


An alternative explanation is suggested by an examination of the Hugoniot curves shown in 
Fig.~\ref{fig:Hugs}. There, in the region of $0.55 < v/v_0 < 0.7$ (which spans the region 
containing the CJ states), there appears to be an approximately linear offset of the curves.  
This suggests that a Mie-Gr{\"u}neisen EOS with an unspecified 
reference Hugoniot might be suitable, where it is assumed that the value of 
$\Gamma /v$ is dependent on $v$. Truncating the Taylor expansion around a reference 
Hugoniot as in \cite{Davis3}, we get 
\begin{equation}\label{eq:EOS}
P\approx P_R + \Gamma (E - E_R )/v ,
\end{equation}
where the Gr{\"u}neisen gamma 
\begin{equation}\label{eq:Gamma}
\Gamma \equiv v \left ( \frac{\partial P}{\partial E} \right ) _v .
\end{equation}
We substitute Eq.~\ref{eq:Hug} into 
Eq.~\ref{eq:EOS} and, because we take $E$ as relative to the products, set 
$E_0 = q$, where $q$ is the specific heat of 
reaction. Upon rearrangement we get for the $P$-$v$ Hugoniot 
\begin{equation}\label{eq:P_Hug}
P_H = \frac{1}{v\left(1-\frac{\Gamma}{2}\frac{v_0 -v}{v}\right)}
\left[v P_R + \Gamma (q-E_R)\right].
\end{equation}
Rearranging Eqs.~\ref{eq:cons_mass} and \ref{eq:cons_momentum}, we find
\begin{equation}\label{eq:cons_mass_mom}
u_s^2=P_H \frac{v_0}{1-\frac{v}{v_0}}.
\end{equation}
This yields the general expression
\begin{equation}\label{eq:gen_us2}
u_s^2 = A(v,\Gamma)P_R + B(v,\Gamma)(q-E_R).
\end{equation}
It should be noted that $q$ does not necessarily equal $Q/m$. In the case where the bond 
order parameter $\overline{B_{ij}}=1.0$ for the initial state, $Q/m$ should be $q$ less the 
small contribution of the van der Waals interaction. When we run an NVE simulation of 1250 AB 
molecules at the initial state used in the NEMD simulations, we find a total internal 
energy of -2548.5 eV\@. Where Davis measures the zero of energy as cold products 
\cite{Davis3}, our calculations use a zero of cold dissociated atoms. 
Our measurement here finds 
$E_0 \approx D_e^{\textrm{AB}}$ and is consistent with $\overline{B_{ij}}=1.0$ and 
a 2\% van der Waals contribution of a few neighbors, therefore $Q/m \approx q$ to very good 
order, and the two can be used interchangeably.

\begin{figure}
\includegraphics*[width=\columnwidth]{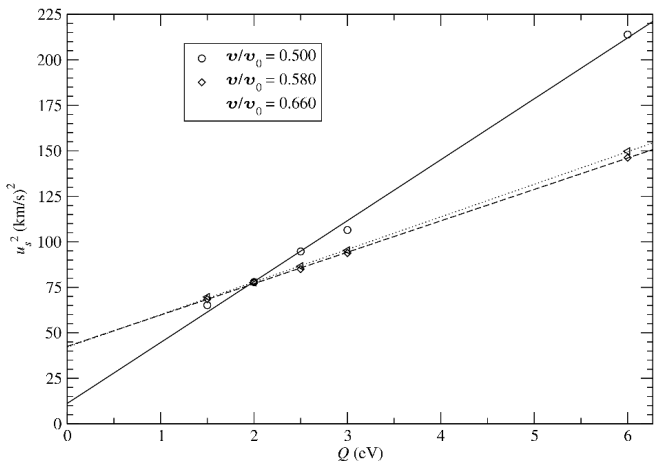}
\caption{Square of the shock velocity ($u_s^2$) vs.~exothermicity ($Q$) 
for select constant values of specific volume ($v$). The lines are linear fits. Notice 
that the fit is better for the higher values of $v$. Notice, also, that the lines 
cross at $v/v_0=2.0$~eV\@. This indicates a linear dependence between the slopes and 
the y-intercepts.
\label{fig:us2vQ_v}}
\end{figure}

\begin{figure}
\includegraphics*[width=\columnwidth]{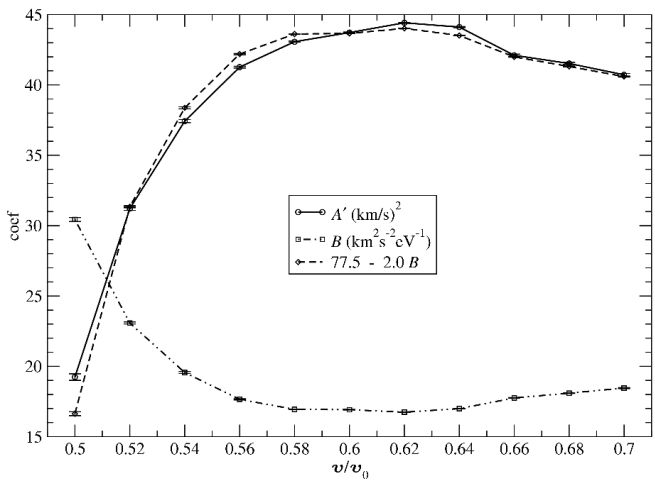}
\caption{Coefficients of the fits as described in Fig.~\ref{fig:us2vQ_v} vs.~specific volume 
for $D_e^{\textrm{AB}}=2.0$~eV. 
$A'$ is the y-intercept and, as in Eq.~\ref{eq:gen_us2}, 
$B$ is the slope. The dashed line is the calculation shown in the legend.
\label{fig:coef_us2vQ_v}}
\end{figure}

The validity of this Mie-Gr{\"u}neisen approximation can be tested by the linear dependence 
of $u_s^2$ on $Q$ for constant volumes.  A few of these plots are given in
Fig.~\ref{fig:us2vQ_v}, which shows that good linear relationships are found for the three 
selected volumes.  This shows that the Mie-Gr{\"u}neisen EOS is a good approximation for 
our model for low $v$ (high compressions). Since the data were generated during the process 
of seeking the minimum $u_s$ for each $Q$, interpolation was used to find values at common 
abscissas. The coefficients of the fits generated by the method in Fig.~\ref{fig:us2vQ_v} are 
then plotted in Fig.~\ref{fig:coef_us2vQ_v} vs. $v/v_0$. Here, $A'$ is the y-intercept of 
those linear fits, and $B$ is the slope. The value $A'$ is distinct from $A$ in 
Eq.~\ref{eq:gen_us2} because it now includes a contribution from $E_R$.  However, the value 
of $B$ is exactly the same as defined in that equation. The fact that all of the lines in 
Fig.~\ref{fig:us2vQ_v} cross at $Q=2.0$~eV leads one to conclude that $A'$ and $B$ are 
linearly dependent. The third curve on Fig.~\ref{fig:coef_us2vQ_v} is an attempt to show 
this for the relation $A' = 77.5 - 2.0 B$.  

The net result of this analysis is that, over the range of $0.55 < v/v_0 < 0.7$, $B$ is 
reasonably constant with a value of 17--19~$(\textrm{km/s})^2 /$~eV\@.  
This result, inserted back into 
Eq.~\ref{eq:gen_us2}, is then in good agreement with the initial observation shown in 
Fig.~\ref{fig:vel2bth}, 
that the slope of the dependence of $u_s^2$ on $Q$ is 
$\approx 19 \textrm{~(km/s)}^2 / \textrm{eV}$\@.  
The ability to approximate the product EOSs as linear Mie-Gr{\"u}neisen offsets of one 
another, at least in the region near the CJ states, clarifies the origin of this linear 
dependence for this system. 

\begin{figure}
\includegraphics*[width=\columnwidth]{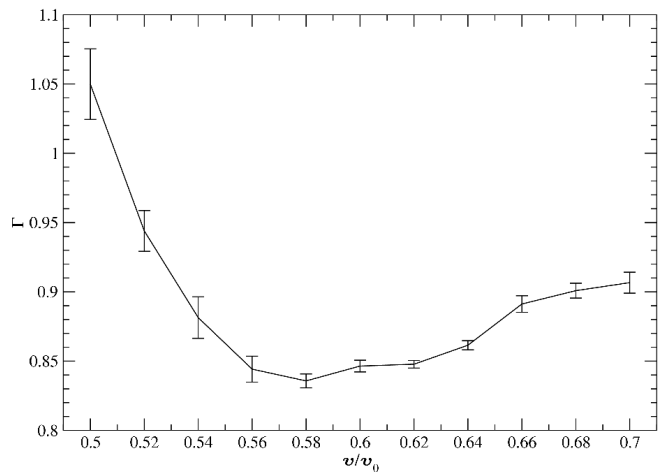}
\caption{Gr{\"u}neisen $\Gamma$ vs.~specific volume for $D_e^{\textrm{AB}}=2.0$~eV. 
\label{fig:Gamma}}
\end{figure}

Using Eqs.~\ref{eq:P_Hug}, \ref{eq:cons_mass_mom}, and \ref{eq:gen_us2}, we can then solve for 
$\Gamma$ to find 
\begin{equation}\label{eq:Gammavv}
\Gamma (v)=\frac{2 B\left(1-\frac{v}{v_0}\right)\frac{v}{v_0}}
{2+B\left(1-\frac{v}{v_0}\right)^2}
\end{equation}
and plot the result in Fig.~\ref{fig:Gamma}. Solving for $B$ we have 
\begin{equation}\label{eq:Bvv}
B(v) = \frac{\Gamma}{\left(1-\frac{v}{v_0 }\right)\left(1-\frac{\Gamma}{2}
\left(1-\frac{v}{v_0 }\right)\right)}.
\end{equation}
Note that $(1-v/v_0 )$ is the compression ($\eta$).  From these equations, it can be seen 
that a constant value of $B$ does not specify a constant value of $\Gamma$, but rather a 
particular dependence of $\Gamma$ on $v$.  Conversely, a constant value of $\Gamma$ would 
also specify a particular dependence of $B$ on $v$.  However, from these two figures it 
is observed that both $B$ and $\Gamma$ are rather weak functions of $v$ in the CJ region.  
This aspect of the EOS behavior (which we are unable to directly adjust) is the origin of 
the linear dependences observed in Fig.~\ref{fig:vel2bth}.


\section{Reaction Zone Thickness and Rate Stick Failure Diameter}
\label{sec:crit}
Having established the steady-state properties of these systems, we now turn to the 
transient properties related to the reaction zone thickness.  There are two means that 
could be employed to determine the character of the reaction zone for these simulations.  
The most direct means is to drive the system with a piston whose velocity is matched to 
that of the CJ conditions.  In this case, the one dimensional profile should exhibit a rapid 
shock rise up to the von Neumann state, followed by a relaxation to the CJ state, which is 
then a constant zone that extends back to the piston.  There will be some transients 
present because the initiation process incurs a slight delay so that an equilibrium state 
is not immediately established, but otherwise this is a 
direct method if the CJ conditions are known.  Less direct is to study a system with either 
an underdriven piston or perhaps initiated with a flyer plate.  Such systems will initially 
propagate at less than the CJ conditions because the release waves can eat into the 
reaction zone.  As the detonation proceeds and the release Taylor wave becomes more spread 
out (approximating a more steady condition behind the reaction zone), these will eventually 
``build-up'' to a steady detonation.  It can be difficult to determine exactly when the 
system has evolved to a steady-state condition.  

\begin{figure}
\raggedright
(a)\\
\includegraphics*[width=\columnwidth]{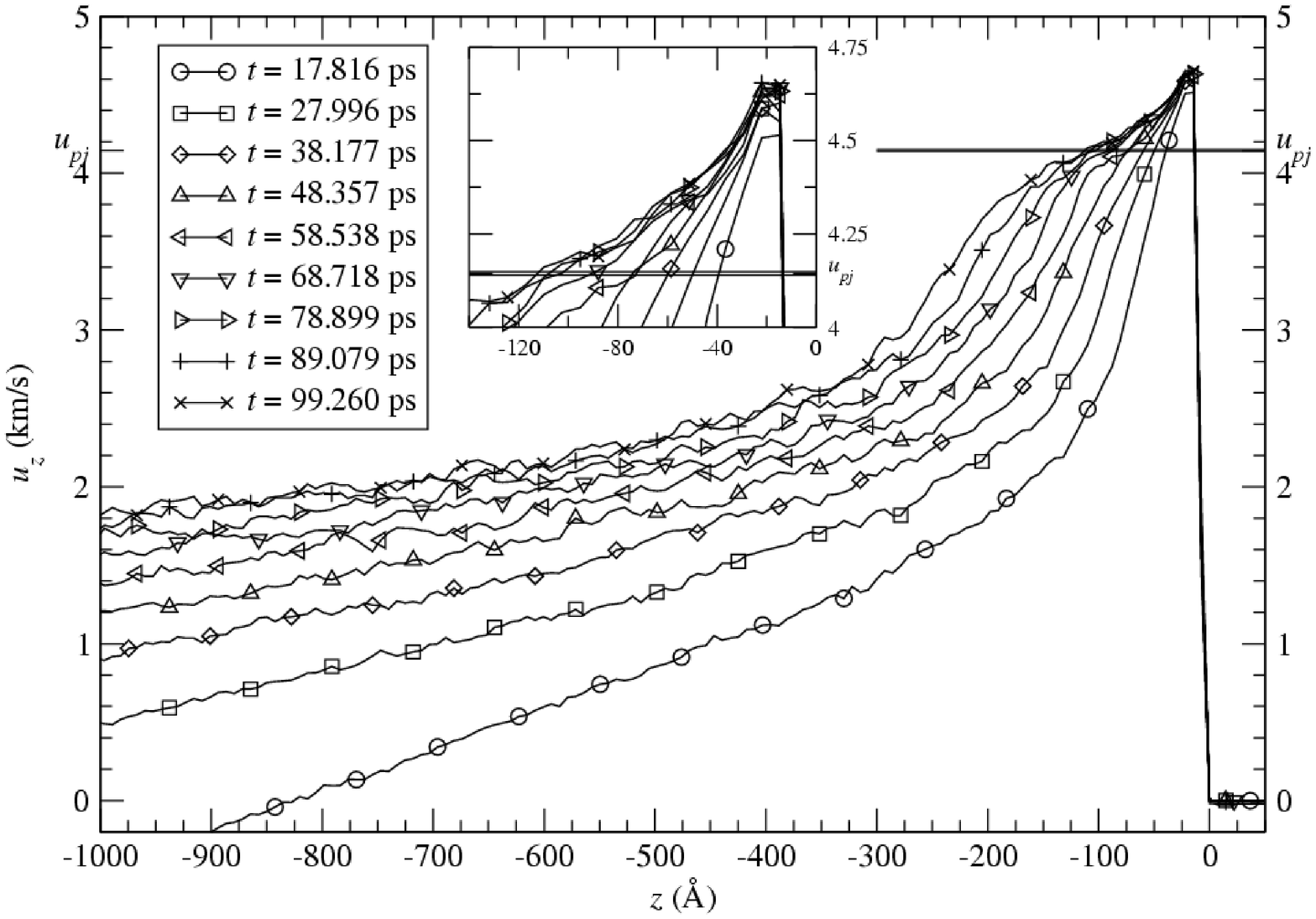}\\
(b)\\
\includegraphics*[width=\columnwidth]{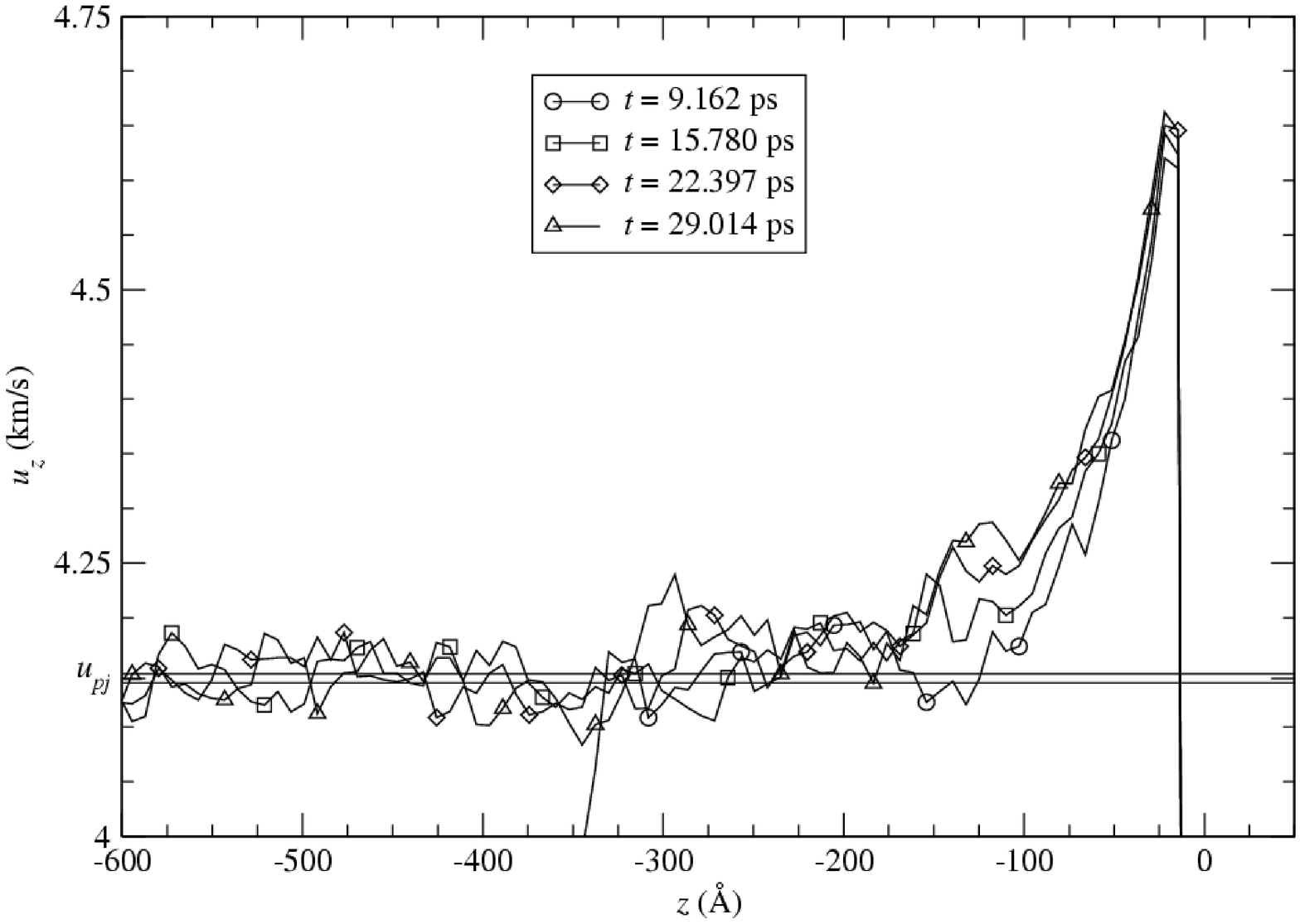}
\caption{(a) 129 profiles of the particle velocity on the z direction ($u_z$) separated in 
time by 50.9 fs are overlapped onto their center profile (whose time is indicated in the 
legend) such that their fronts line up. 
They are averaged and those averages are similarly overlapped. The simulation represented has 
the default parameterization for the REBO potential \cite{Brenner}. It is a flyer driven 
unsupported simulation as described in Sec.~\ref{sec:method}. Found for this specific 
parameterization in Sec.~\ref{sec:eos}, $u_{pj}$ is indicated by a constant line segment. 
The inset is a magnification of the region containing the von Neumann spike and the CJ state.\\
(b) Similar profiles for a critically supported detonation. This simulation is shorter and 
thinner than in (a), but it suggests that in (a) still has some building to do for the 
profiles do not seem to settle down to $u_{pj}$ until about 300~\AA~behind the front.
\label{fig:ovrlpdavgvz}}
\end{figure}

A comparison of these two approaches is shown in Fig.~\ref{fig:ovrlpdavgvz}.  For the case 
of a piston matched to the CJ conditions, a steady solution evolved quite rapidly as 
determined by both the detonation velocity and the constant properties of the following 
zone.  The reaction zone, defined to be the distance from the shock front to the point 
where the transient properties are not distinguishable from the thermal fluctuations of the 
constant zone, extends out to $\approx 300$~\AA.  A somewhat better characterization 
is probably the 
distance at which the particle velocity has decreased to $1/e$ of the initial overshoot.  This 
occurs at $\approx 60$~\AA~behind the shock front, or a characteristic time constant of 
0.7~ps.  For 
the transient simulations, it is apparent that the system is continuing to evolve even after 
propagating for 100 ps.  The reaction zone length (emphasized in the inset) is clearly 
extending past 100~\AA.  (The unusual structure after the reaction zone is likely due to the 
unusual chemical structure of the CJ state, and could be characterized as a phase transition 
back to the diatomic $\textrm{A}_2$ and $\textrm{B}_2$ products.)  
These measurements for the build up time, reaction zone width, and the detonation 
velocity are larger than those proposed by White \emph{et al.} \cite{White} for the same model. 
They include in their calculations a time domain in which the detonation is still subtly 
building. We, in fact, measure a detonation velocity ($9.5580 \pm 0.0013$~km/s) close to 
theirs (9.5~km/s) for the unsupported case if we include the front positions at 
10~ps~$< t < 20$~ps in the linear least-squares fitting that determines $u_s$.  
This highlights the difficulties in using this approach of unsupported detonations. 

It should be noted that the transient system exhibits the classic build-up behavior quite 
nicely.  In addition to the extension of the reaction zone out to its equilibrium value, 
there is also the gradual increase in the initial shock state up to its equilibrium value.  
As this sets the initial temperature and rate for the reaction chemistry, it emphasizes 
the significance of these transient phenomena.  The shock velocity is also a sensitive 
indicator of the build-up process.  While the deprecation of the velocity below the 
steady-state value is not great nor readily obvious, it can be discerned by careful 
observation.  Previously reported discrepancies between expected and observed CJ 
performance may well have been due to systems not having fully attained steady-state 
conditions. 

In \cite{Davis2} Davis shows that surface effects adversely affect detonation's
ability to propagate because rarefaction waves from the side release erode the reaction 
zone and its support structure. For cylindrical charges, this is characterized as a 
failure diameter below which
detonation cannot propagate. White \emph{et al.} report that, when the periodic
boundaries are removed from the sample so that the particles are free to 
escape the system, there is a critical width ($W_{c}$) that the sample must 
exceed in $\hat{x}$, the direction perpendicular to the propagation of the 
shock front, in order that the detonation be sustained. If the sample is too 
thin, rarefaction waves quench the detonation 
\cite{White}.

\begin{figure}
\includegraphics*[width=\columnwidth]{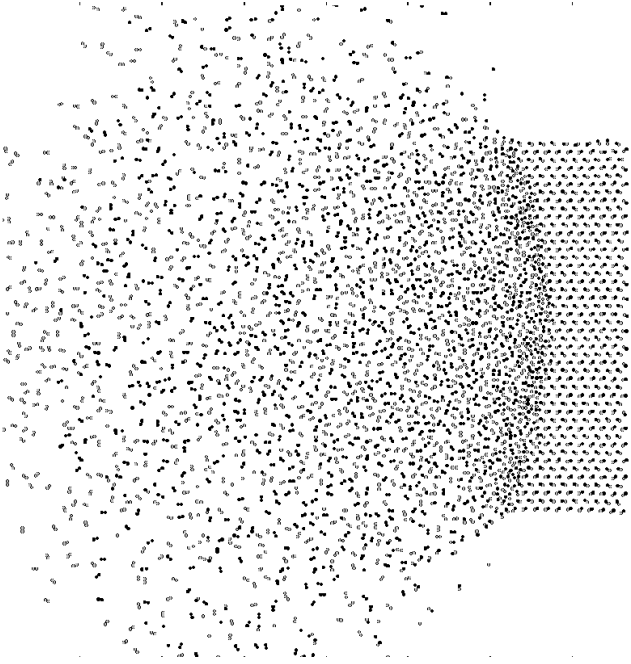}
\caption{Rate stick at super-critical width.
\label{fig:ratestick}}
\end{figure}

\begin{figure}
\includegraphics*[width=\columnwidth]{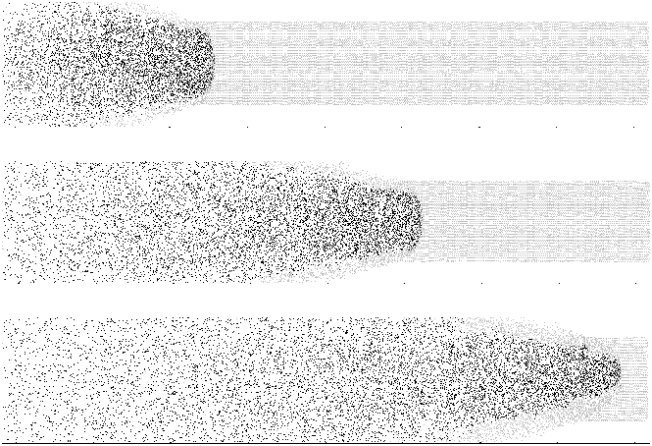}
\caption{Three sequential snapshots of a rate stick at sub-critical width. 
Here the particles are colored by bond type. Gray is unreacted, and black is reacted.
Rarefaction waves from the rate stick's edge erode at the reaction zone thus quenching the 
detonation.
\label{fig:failstick}}
\end{figure}

\begin{figure}
\includegraphics*[width=\columnwidth]{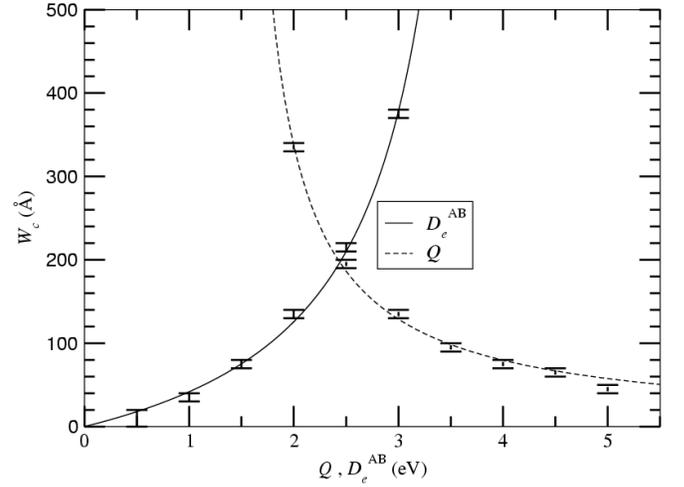}
\caption{Critical width ($W_{c}$) vs.~AB dissociation energy ($D_{e}^{\textrm{AB}}$) and 
exothermicity ($Q$). The upper dash of the error bar indicates the thinnest 
sample to sustain detonation. The lower error bar indicates the thickest 
sample to fail within 360~lattice~spaces. The resolution is 10~\AA\@.
The lines are guides to the eye with empirical functional forms discussed in 
the text.\label{fig:cwbth}}
\end{figure}

$W_{c}$'s dependence on $Q$ \cite{Fellows} and $D_{e}^{\textrm{AB}}$ can be seen in  
Fig.~\ref{fig:cwbth}. Unlike in \cite{Fellows}, a curve of the form 
$y=a/(x-Q_c )$, not exponential, is fit to the data for $Q$. Simpler 
equations fit better, but this form is inspired by Hubbard and Johnson's \cite{Hubbard} 
dependence of delay time ($t_{d}$) (described below) on $Q$ and to be asymptotic at 
the point where detonation is known to fail for an infinitely thick sample ($Q_c$). 
For $D_{e}^{\textrm{AB}}$ a curve of the form $y=a(e^{bx}-1)/(x-D_{ec}^{\textrm{AB}})$ 
is fit to the data for similar reasons. 
If $Q$ is raised, the reaction should be more difficult to quench because 
there is more energy available to break neighboring AB bonds. The reaction is expected to 
be faster and the reaction zone shorter because of the higher temperature  resulting from 
the greater energy release. As $Q$ lowers 
toward $Q_{c}$, no matter how wide the sample is, detonation will not be sustained. On the 
other hand, if $D_{e}^{\textrm{AB}}$ is lowered, it is harder for 
rarefaction waves to quench the 
detonation front because, with a lower dissociation energy, a chemical reaction 
occurs more readily. $W_{c}$ is, thus, lowered because 
detonation is more easily sustained. As it rises, $D_{e}^{\textrm{AB}}$ reaches a critical 
value ($D_{ec}^{\textrm{AB}}$) (probably dependent on $Q$, flyer thickness and velocity, 
etc.) above which detonation cannot be sustained, no matter how wide the sample.

In a sample with culled boundaries, if $D_{e}^{\textrm{AB}}$ is lowered, it is harder for 
rarefaction waves to quench the 
detonation front because, with a lower dissociation energy, a chemical reaction 
occurs more readily. $W_{c}$ is, thus, lowered because 
detonation is more easily sustained. As it rises, $D_{e}^{\textrm{AB}}$ reaches a critical 
value ($D_{ec}^{\textrm{AB}}$) (probably dependent on $Q$, flyer thickness and velocity, 
etc.) above which detonation cannot be sustained, no matter how wide the sample.

Using the idea of delay time ($t_{d}$), during which a reactant must remain 
above a certain temperature in order for a reaction to occur, Hubbard 
\emph{et al.} give an example of how a detonation's ability to be initiated 
or sustained may depend on 
$Q$ and $D_{e}^{\textrm{AB}}$. We use this in lieu of finding reaction rate data. 
The longer this time, the less likely the reaction. In their 
example, $t_{d}$ has an inverse dependence on $Q$ and an inverse and exponential dependence 
on activation energy, $E_{a}$ (Eq.~\ref{eq:delay}) \cite{Hubbard}.
\begin{equation} \label{eq:delay}
t_{d}=\frac{R E^{2}_{0}}{c_{v} \nu E_{a} Q} \exp \left( \frac{c_{v} E_{a}}{R E_{0}} 
\right),
\end{equation}
where $\nu$ is the collision rate, $R$ is the molar gas constant, $c_{v}$ is 
the specific heat at constant volume, and $E_{0}$ is the internal energy behind
the shock.

In Rosing and Chariton theory, when $\Theta=t_{r}$, 
$d=W_{c}$, where $\Theta$ is scattering time, $t_{r}$ is the reaction time, 
and $d$ is the diameter of the sample. 
The rate stick must be thick enough that rarefaction waves, traveling at 
$c$ from the 
surface, cannot sufficiently penetrate the HE in order to scatter the reaction zone and 
quench the reaction in a time equal to the reaction time. 
$a=t_{r}(u_{s}-\bar{u})$, where $a$ is the width of the chemical reaction zone, 
$\bar{u}$ is the average particle velocity in $\hat{z}$ within the zone, and 
$u_{s}$ is the velocity of the detonation front. $\Theta=d/2c$, where $c$ is the 
speed of sound. Replacing $t_{r}$ with $\Theta$, one finds 
\begin{equation}\label{eq:Wcva}
W_{c}=d=2ca/(u_{s}-\bar{u}),
\end{equation}
where $u_{s}$ is a function of $d$ 
\cite{Dremin,Swanson,Davis2}\@. 
If $t_{d} \propto t_{r}$ and $E_{a} \propto D_{e}^{\textrm{AB}}$, we find that  
$W_{c}\propto t_{d}$, supporting our selection of curve fits.

To check the validity of Eq.~\ref{eq:Wcva}, we use for $a$ the characteristic decay length of 
the reaction zone (60~{\AA}) determined above.
If the ratio $2c/(u_s -\bar{u}) \sim 2$, this would be excellent agreement.  As the value 
of $u_s -\bar{u}$ should be approximately equal to the local sound speed, this estimation 
is then quite good.  Given the approximations involved in making this 
assessment, this is certainly a fortuitous agreement, but can be taken as support for this 
analysis.  Overall, a more thorough understanding of these phenomena is evolving through 
these studies.

\section{Conclusion}

We have analyzed the relationships between the energetic
chemical properties of a simple, but well studied, model high
explosive and the physical properties of 
its detonation front through explicit Molecular Dynamics (MD).  This provides an excellent 
means to compare to continuum analyses and avoids the
complication of a multiphase equation of state (EOS) and other approximations required 
in those approaches. Most previous MD studies
of high explosives have focused on the physical properties of the HE as
independent parameters (e.g.~voids and gaps) to validate this general approach. 
Our analysis has been conducted in order to establish a more quantitative comparison to
the relevant theories by, e.g., seeking the CJ state via NVE
simulations and showing that it accurately predicts detonation properties such as 
the detonation velocity ($u_s$). Despite spurious model anomalies, e.g., a CJ state 
consisting of polyatomic species and unusual properties of the product Hugoniot curves, 
we have observed general consistency between the MD approach and the thermodynamic
equations.  Given that the fundamental CJ and ZND models pose minimal constraints on the 
behavior of chemical system, this result is not extremely surprising.  However, we have 
demonstrated that it is possible to gain a quantitative understanding of these 
relationships even for these somewhat unusual systems. For the analysis of the 
behavior of $u_s$ and the critical width as functions 
of energetic parameters,
we have explained our results in the context of available theories. In the
case of $u_s^2$ as a function of the dissociation energy of AB, we have not 
directly determined the reason for its linear 
and increasing relationship, but assume that it also determined by the induced EOS changes. 
In characterizing the reaction zone width, we demonstrated 
that transient simulations must be scrutinized rather carefully in
determining the final state.
We are currently investigating the sources of the model anomalies that we have identified, 
and potential model revisions that will correct these are expected to result in a system 
that behaves more similarly to conventional high explosives.

\acknowledgments
The authors would like to thank Sam Shaw, Alejandro Strachan, Jerome Erpenbeck, 
Thomas Sewell, Shirish Chitanvis, Tamer Zaki, Yogesh Joglekar, Ben Haley, 
Rolando Somma, and David Hall for useful conversations. This material was 
prepared by the University of California under Contract W-7405-ENG-36 with the 
U.S. Department of Energy. The authors particularly wish to recognize funding 
provided through the ASC Physics and Engineering Modeling program.

\bibliography{papbib}
\end{document}